\documentclass[%
 reprint,
 amsmath,amssymb,
 aps,
]{revtex4-1}

\usepackage{graphicx}% Include figure files
\usepackage{dcolumn}% Align table columns on decimal point
\usepackage{bm}% bold math
\newcommand{\be}{\begin{equation}}
\newcommand{\ee}{\end{equation}}
\newcommand{\bea}{\begin{eqnarray}}
\newcommand{\eea}{\end{eqnarray}}
\newcommand{\nn}{\nonumber}
\begin{document}

\preprint{APS/123-QED}

\title{Light quark pseudoscalar  densities and anomaly matrix elements for $\eta$ and $\eta'$
mesons}

\author{Janardan P. Singh}
 \affiliation{  Physics Department, Faculty of Science, The M. S. University of Baroda, Vadodara-390002, India.}
 \email{janardanmsu@yahoo.com}
\date{\today}% It is always \today, today,

\begin{abstract}
Matrix elements of flavor-diagonal light quark pseudoscalar densities and axial anomaly operators between vacuum and $\eta$ and $\eta'$ meson states have been determined. This has been done by evaluating current-current correlators of octet-octet  and octet-singlet axial currents using QCD sum rules. The numerical values obtained for the matrix elements compare well with those obtained in the current literature.

\begin{description}
\item[PACS numbers]
11.15.Tk, 12.38.Lg, 11.40.Ex, 14.40.Be
\end{description}
\end{abstract}

\maketitle

%\tableofcontents

\section{INTRODUCTION}
Matrix elements of pseudoscalar density operators between vacuum and $\eta$ and $\eta'$ mesons are useful in study of processes involving production and decay of these mesons. In particular, they play an important role in semi-leptonic and nonleptonic  B meson decays involving $\eta$ and $\eta'$ mesons [1-8]. These as well as the anomaly matrix elements are useful in discussion of pseudoscalar glueballs [9]. The latter have also been used in study of gluonic components of $\eta$- $\eta'$ mesons [10]. These matrix elements also appear in the discussion of two-parton light-cone distribution functions of these mesons where there is no direct relation between coupling constants of twist-three operators with those of twist -two operators for $\eta$ and $\eta'$ mesons [11]. In this work we will determine the matrix elements
\be
    \langle 0 |m_{q}\bar{q}i\gamma_{5}q|\eta, \eta'\rangle,\hspace{4pt} q=u,d,s
\ee
and

\be
    \langle 0 |\frac{\alpha_{s}}{4\pi}G^{a}_{\mu\nu}\widetilde{G}^{a\mu\nu}|\eta, \eta'\rangle
\ee
using QCD sum rule. For this, we first consider the correlators of two axial vector currents
\be
    \Pi^{ab}_{\mu\nu}=i \int d^{4}x \; e^{iqx}\langle 0|T\{ J^{a}_{\mu 5}(x),J^{b}_{\nu 5}(0)\}|0 \rangle ;
    \;\;(a,b=8,0)
\ee
where
\bea
J^{8}_{\mu 5}=\frac{1}{\sqrt{6}}(\overline{u}\gamma_{\mu}\gamma_{5}u+\overline{d}\gamma_{\mu}\gamma_{5}d-2\overline{s}\gamma_{\mu}\gamma_{5}s)
\nonumber\\
J^{0}_{\mu 5}=\frac{1}{\sqrt{3}}(\overline{u}\gamma_{\mu}\gamma_{5}u+\overline{d}\gamma_{\mu}\gamma_{5}d+\overline{s}\gamma_{\mu}\gamma_{5}s)
\eea

We had earlier considered the correlators (3) in Ref. [12] and determined the decay constants of $\eta$ and $\eta'$ mesons for the above octet and singlet currents defined by
\bea
    \langle 0 |J^{a}_{\mu 5}|P(p) \rangle=if^{a}_{P}p_{\mu};\;\;P=\eta, \eta'
\eea
\vspace{2pt}

\noindent in the limit $m_{u}=m_{d}=0 $. In this work, we determine their values keeping all the quark masses to be nonzero. In the current literature the decay constants $f^{a}_{P} $ are parameterized as
\be
    \left( \begin{array}{clrr} %
        f^{8}_{\eta} & f^{0}_{\eta}  \\
        f^{8}_{\eta'}  & f^{0}_{\eta'}
    \end{array} \right)
    = \left( \begin{array}{clrr} %
        f_{8}\cos\theta_{8} & -f_{0}\sin \theta_{0} \\
        f_{8}\sin \theta_{8}  & f_{0}\cos\theta_{0}
    \end{array} \right)
\ee

On forming the divergences of polarization tensor $\Pi^{ab}_{\mu\nu}(q)$ with the momentum, we get
\be
q^{\mu} \Pi^{ab}_{\mu\nu}(q)q^{\nu}=-P^{ab}_{L}(q^{2})q^{2}
\ee
where $P^{ab}_{L}(q^{2})$ is free from kinematic singularities. The divergences of currents are given as
\bea
\partial^{\mu}J^{8}_{\mu 5}=\frac{2i}{\sqrt{6}}(m_{u}\overline{u}\gamma_{5}u+m_{d}\overline{d}\gamma_{5}d-2m_{s}\overline{s}\gamma_{5}s)
\mbox{\hspace{70pt}}\nonumber\\
\partial^{\mu}J^{0}_{\mu 5}=\frac{2i}{\sqrt{3}}(m_{u}\overline{u}\gamma_{5}u+m_{d}\overline{d}\gamma_{5}d+m_{s}\overline{s}\gamma_{5}s)
\mbox{\hspace{73pt}}\nonumber\\
-\frac{\sqrt{3}\alpha_{s}}{4\pi}G^{a}_{\mu\nu}\widetilde{G}^{a\mu\nu}\mbox{\hspace{80pt}}\nonumber\\
\eea
where we have defined dual field strength tensor as
\be
\widetilde{G}^{a\mu\nu}=\frac{1}{2}\varepsilon^{\mu\nu\rho\sigma}G^{a}_{\rho\sigma},\;\;\varepsilon^{0123}=+1
\ee
We will assume isospin symmetry in the quark matrix elements:
\bea
\langle0|m_{u}\overline{u}i\gamma_{5}u|\eta\rangle=\langle0|m_{d}\overline{d}i\gamma_{5}d|\eta\rangle \equiv  m_{q} A_{q}\\
\langle0|m_{u}\overline{u}i\gamma_{5}u|\eta'\rangle=\langle0|m_{d}\overline{d}i\gamma_{5}d|\eta'\rangle \equiv  m_{q} A'_{q}
\eea
and call
\bea
 \langle0|m_{s}\overline{s}i\gamma_{5}s|\eta\rangle \equiv  m_{s} A_{s}\\
 \langle0|m_{s}\overline{s}i\gamma_{5}s|\eta'\rangle \equiv  m_{s} A'_{s}
\eea
while the anomaly matrix elements will be denoted as
\bea
\langle 0 |\frac{\alpha_{s}}{4\pi}G^{a}_{\mu\nu}\widetilde{G}^{a\mu\nu}|\eta\rangle\equiv A_{G}\\
\langle 0 |\frac{\alpha_{s}}{4\pi}G^{a}_{\mu\nu}\widetilde{G}^{a\mu\nu}|\eta'\rangle\equiv A'_{G}
\eea
In Eqs. (10) and (11), $ m_{q} $ is the average of u- and d-quark masses. The strange-quark matrix elements $ A_{s} $ and $ A'_{s} $ will be determined in the limit  $ m_{q} $=0. For this, we require the values of meson masses $ m_{\eta} $ and $ m_{\eta'} $ also in the same limit. From the diagonalization of $ m_{\eta}- m_{\eta'} $ mass matrix [13], one obtains values for $\eta $ and $\eta' $ masses:
\be
m^{2}_{\eta',\eta}=(m^{2}_{K}+\widetilde{m}^{2}_{\eta_{0}}/2 )\pm\frac{1}{2}\sqrt{(2m^{2}_{K}-2m^{2}_{\pi}-\frac{1}{3}\widetilde{m}^{2}_{\eta_{0}} )^{2}+\frac{8}{9}\widetilde{m}^{4}_{\eta_{0}}}
\ee
Here $ \widetilde{m}^{2}_{\eta_{0}}	$ is the gluonic mass term having a rigorous interpretation through the Witten-Veneziano mass formula [14,15]. On taking the value $ \widetilde{m}^{2}_{\eta_{0}} $=0.73 $ GeV^{2} $, Eq. (16) agrees with the physical masses within 10\% range. The dependence of $ m_{K} $ and $ m_{\pi} $  on light quark masses are well known [16]. From this, we find that  $ m_{\eta} $ decreases by 2.2\% while $ m_{\eta'} $ decreases by 0.65\% on setting $ m_{q} $ =0. Using this result for the experimental masses $ m_{\eta} $=0.547 GeV and $ m_{\eta'} $=0.958 GeV, we find that the values of $\eta $ and $\eta' $ masses are
\be
\widetilde{m}_{\eta}=0.535\; GeV,\;\;\widetilde{m}_{\eta'}=0.952 \; GeV
\ee
in a world where  $ m_{q} $=0.
 We will assume $\eta $ and $\eta' $ to be made up of only light quarks with no mixing with $c\bar{c} $ system or any glueball state. Among half a dozen functions considered in Ref.[12], we will consider here sum rules only for $P^{88}_{L} $ and $P^{08}_{L} $. Sum rules for these functions were independent of any instanton contribution and the quality of the fit of the phenomenological  side with the OPE side was best obtained for these functions. In addition to this, these sum rules give us sufficient information for our purpose.\\

     To use the method of QCD sum rule for this purpose, as is well known, one calculates $P^{ab}_{L} $ using OPE on the one hand and it is evaluated phenomenologically using dispersion integral on the other hand. Borel transform of the two sides are matched over an appropriately chosen Borel window. After Borel transform, phenomenological side is dominated by the ground state contribution while the resonance  and the continuum contributions are parameterized by the loop diagrams of the OPE side with a continuum threshold. Matching of the two sides determines the phenomenological quantity of interest in terms of QCD parameters which includes vacuum condensates.\vspace{-1cm}
\section{SUM RULES}
From $P^{88}_{L} $, we get the following sum rule which is an extension of the corresponding Eq. (33 ) in Ref. [12] with $ m_{q}\neq0 $ as
%------ Edit Equation ---------------
\bea
    K_{88} + \frac{8}{3}\frac{1}{m^{2}_{\eta}}(m_{q}A_{q}- m_{s}A_{s})^{2}e^{-m^{2}_{\eta}/M^{2}}
                \mbox{\hspace{80pt}} \nonumber\\
           + \frac{8}{3}\frac{1}{m^{2}_{\eta'}}(m_{q}A'_{q}- m_{s}A'_{s})^{2}e^{-m^{2}_{\eta'}/M^{2}}
                \mbox{\hspace{76pt}} \nonumber\\
    =\frac{1}{\pi^{^{2}}}m_{s}^{2}M^{2}\left\{1+\frac{\alpha_{s}}{\pi}\left(\frac{17}{3}+2\gamma-2ln\frac{M^{2}}{\mu^{2}} \right) \right\}E_{0}\left(\frac{W^{2}}{M^{2}}\right)
                \mbox{\hspace{10pt}}\nonumber\\
        -\frac{8}{3}m_{s}\langle\bar{s}s\rangle+\frac{16}{3}m^{3}_{s}\frac{\langle\bar{s}s\rangle}{M^{2}}-
         \frac{2}{3}\frac{m^{2}_{s}}{M^{2}} \left \langle\frac{\alpha_{s}}{\pi}G^{2}\right\rangle
                \mbox{\hspace{47pt}}\nonumber\\
        -\frac{64}{9}\pi^{2}\frac{\alpha_{s}}{\pi}\frac{m^{2}_{s}}{M^{4}}\kappa_{s}\langle\bar{s}s\rangle^{2}-\frac{4}{3}
         \frac{m^{3}_{s}}{M^{4}}\langle\bar{s}g_{s}\sigma.G s\rangle-\frac{4}{3}m_{q}\langle\bar{q}q\rangle \mbox{\hspace{-2pt}}\nonumber\\
        +\frac{1}{2\pi^{^{2}}}m_{q}^{2}M^{2}\left\{1+\frac{\alpha_{s}}{\pi}\left(\frac{17}{3}+2\gamma-2ln\frac{M^{2}}{\mu^{2}} \right) \right\}E_{0}\left(\frac{W^{2}}{M^{2}}\right)\mbox{\hspace{2pt}}\nonumber\\
        -\frac{1}{3}\frac{m^{2}_{q}}{M^{2}}  \left\langle\frac{\alpha_{s}}{\pi}G^{2}\right\rangle-\frac{32}{9}\pi^{2}\frac{\alpha_{s}}{\pi}\frac{m^{2}_{q}}{M^{4}}
        \kappa\langle\bar{q}q\rangle^{2} \mbox{\hspace{62pt}}
  \eea
%------------------------------------
In the above equation, M is the Borel mass parameter, $\mu $ is the renormalization point,$ \kappa $ and $ \kappa_{s}$ are different from 1 to account for contributions from beyond ground-state factorization of four-quark condensate [17], not considered in Ref. [12],\;$ \gamma $ is the Euler's constant, W is the continuum threshold,\;$ E_{0} (x)=1-e^{-x} $ and $ \langle\bar{u}u\rangle $=$\langle\bar{d}d\rangle $=$\langle\bar{q}q\rangle $. As explained in Ref. [12],\;$ K_{88} $ is the residue of a spurious pole in $P_{L}^{88} $ at $q^{2}=0 $ which is introduced due to approximate form of $ P_{L}^{88} $ obtained by OPE. For u- and d- quark contributions to the RHS of the sum rule we have retained terms up to quadratic in  $ m_{q} $, since the linear term will contribute only to $K_{88} $. Also, clearly we have  relations such as,\;$ \frac{8}{3}  \frac{1}{m_\eta^2} (m_{q} A_{q}-m_{s} A_{s} )^{2}=m_{\eta}^{2} (f_{\eta}^{8})^{2} $, etc. [12]. On putting $ m_{q}=0 $, we get the sum rule
\bea
\widetilde{K}_{88}+\frac{8}{3}\frac{1}{\widetilde{m}^{2}_{\eta}}(m_{s}A_{s})^{2}e^{-\widetilde{m}^{2}_{\eta}/M^{2}} \mbox{\hspace{120pt}}\nonumber\\+\frac{8}{3}\frac{1}{\widetilde{m}^{2}_{\eta'}}(m_{s}A'_{s})^{2}e^{-\widetilde{m}^{2}_{\eta'}/M^{2}} \mbox{\hspace{120pt}} \nonumber\\
    =\frac{1}{\pi^{^{2}}}m_{s}^{2}M^{2}\left\{1+\frac{\alpha_{s}}{\pi}\left(\frac{17}{3}+2\gamma-2ln\frac{M^{2}}{\mu^{2}} \right) \right\}E_{0}\left(\frac{W^{2}}{M^{2}}\right)
                \mbox{\hspace{10pt}}\nonumber\\
        -\frac{8}{3}m_{s}\langle\bar{s}s\rangle+\frac{16}{3}m^{3}_{s}\frac{\langle\bar{s}s\rangle}{M^{2}}-
         \frac{2}{3}\frac{m^{2}_{s}}{M^{2}} \left \langle\frac{\alpha_{s}}{\pi}G^{2}\right\rangle
                \mbox{\hspace{50pt}} \nonumber\\
  -\frac{64}{9}\pi^{2}\frac{\alpha_{s}}{\pi}\frac{m^{2}_{s}}{M^{4}}\kappa_{s}\langle\bar{s}s\rangle^{2}-\frac{4}{3}
  \frac{m^{3}_{s}}{M^{4}}\langle\bar{s}g_{s}\sigma.G s\rangle\mbox{\hspace{50pt}}\nn\\
  \eea
  We have used the following constants in Eqs. (18) and (19) [12,18]:
  \bea
\alpha_{s}(1\; GeV^{2} )=0.5,\;a=-(2\pi)^{2}\langle\bar{q}q\rangle=0.55\;GeV^{3},\nn\\ \langle\bar{s}s\rangle=0.8\langle\bar{q}q\rangle,
b=\langle g^{2}_{s}G^{2}\rangle=0.5\; GeV^{4},\nonumber\\ \langle\bar{s}g_{s}\sigma.G s\rangle=m^{2}_{0}\langle\bar{s}s\rangle\;with\; m^{2}_{0}=0.8\; GeV^{2}, \nonumber\\
m_{s}=0.153\;GeV,\;m_{q}=0.005\;GeV,\;\kappa=\kappa_{s}=2.5,\nn\\ \mu=1\;GeV \;and \; W^{2}=2.3\; GeV^{2}\mbox{\hspace{10pt}}
\eea
Eq.(19) was used in Ref. [12] with $\kappa_{s}=1 $ and physical masses of $ \eta $ and $\eta'$.  The range of $ M^{2} $ over which the two sides of a sum rule are matched is decided as follows. The smaller the $ M ^{2}$, the more important are the higher dimensional operators which puts a lower limit on $ M ^{2}$ while the larger is the $ M ^{2}$, the more important are the resonance and continuum states which puts an upper limit on $ M ^{2}$.  We fit the two sides of each of Eqs. (18) and (19) over a range $ 1.0\; GeV^{2} < M ^{2} < 1.7 \;GeV^{2} $. It is observed that in the specified range the operators of highest dimensional included in the OPE are contributing less than 1\% and the resonance and the continuum states, transferred on OPE side and contained in $ E_{0} (W^{2}/M^{2} ) $ in perturbative terms, are contributing less than 25\% to the OPE side result. This is well within the accepted criterion for the standard treatment of QCD sum rules [18]. We have shown the plots of r.h.s. of Eq.(18) and a fit as l.h.s. in Fig. 1. Also shown are r.h.s. of Eq.(19) along with a fit as l.h.s in Fig. 2. Eq.(18) gives slightly changed result for the decay constants:
\bea
f^{8}_{\eta} &=& 159.1\; MeV (165.6\; MeV),\nn\\
f^{8}_{\eta'}&=& -66.4\; MeV (-62.2\; MeV)
\eea
where the quantity in the bracket is from Ref. [12]. This gives us $ f_{8}=172.4 MeV $, $ \theta_{8} $= $-22.6^{\circ}$. If we show the quality of the fit by $ \chi^{2} $, defined as
\be
\chi^{2}=\frac{1}{N}\sum^{N}_{i=1}\frac{[f(x_{i} )-f_{fit}(x_{i} )]^{2}}{[f(x_{i} )+f_{fit}(x_{i} )]^{2}}
\ee
where $f(x_{i})$ stands for the r.h.s. of Eqs. (18) and (19) and $ f_{fit}(x_{i}) $ for the l.h.s. of the corresponding equation, we find that for  N=20, $ \chi=4.7\times10^{-5}$ for both the curves. The result of the fit for Eq. (18) gives:
\bea
\frac{8}{3}\frac{1}{m^{2}_{\eta}}(m_{q}A_{q}- m_{s}A_{s})^{2}=7.58\times10^{-3}\;GeV^{4} \nonumber\\
\frac{8}{3}\frac{1}{m^{2}_{\eta'}}(m_{q}A'_{q}- m_{s}A'_{s})^{2}=4.04\times10^{-3}\;GeV^{4} \nonumber\\
K_{88}=1.41\times10^{-3}\;GeV^{4}
\eea
while that for Eq. (19) gives
\bea
\frac{8}{3}\frac{1}{\widetilde{m}^{2}_{\eta}}( m_{s}A_{s})^{2}=7.52\times10^{-3}\;GeV^{4} \nonumber\\
\frac{8}{3}\frac{1}{\widetilde{m}^{2}_{\eta'}}( m_{s}A'_{s})^{2}=4.22\times10^{-3}\;GeV^{4} \nonumber\\
\widetilde{K}_{88}=1.20\times10^{-3}\;GeV^{4}
\eea
Eq. (19) was obtained from Eq. (18) by setting $ m_{q}=0 $ consistently on both sides of the equation and this includes the theoretical values of $ \widetilde{m}^{2}_{\eta} $ and $\widetilde{m}^{2}_{\eta'} $ as well in the same limit. Hence, Eqs. (23) and (24), obtained from fittings of Eqs. (18) and (19) respectively, are  two self-consistent independent algebraic equations. As is shown below for $ A_{G} $ and $ A'_{G}$(see Eqs. (28) and (30) below) following a similar procedure, taking limit $ m_{q}=0 $ does not change these matrix elements in a significant way.  We solve Eqs. (23) and (24)  to obtain the following results:
\bea
-m_{s}A_{s} &=& 0.02841\; GeV^{3},\mbox{\hspace{60pt}}\nn\\
 m_{s}A'_{s}&=& 0.03786\; GeV^{3},\nn\\
 m_{q}A_{q} &=& 7.45\times10^{-4}\;GeV^{3},\nn\\
 m_{q}A'_{q}&=& 5.53\times10^{-4}\;GeV^{3}.
\eea
Sign of $ m_{s}A_{s} $ is fixed from the requirement that in the limit of $ m_{q}=0 $, $-m_{s}A_{s}= \frac{\sqrt{6}}{4}f_{\eta}^{8}m^{2}_{\eta} $ which is positive. Eq. (24) gives us $f_{8}=175.9\; MeV $ and $ \theta_{8}=-22.8^{\circ}$.\vspace{0.1in}
\mbox{\hspace{10pt}} From $P_{L}^{08} $, we get the following sum rule which is an extension of the corresponding Eq. (39) in Ref.[12] with $ m_{q}\neq0 $:
\bea
K_{08}+\frac{8\sqrt{2}}{3m^{2}_{\eta}}(m_{q}A_{q}-m_{s}A_{s})\times \mbox{\hspace{130pt}}\nn\\
(m_{q}A_{q}+\frac{1}{2}m_{s}A_{s}-\frac{3}{4}A_{G})e^{-m^{2}_{\eta}/M^{2}} \mbox{\hspace{70pt}}\nn\\
+\frac{8\sqrt{2}}{3m^{2}_{\eta'}}(m_{q}A'_{q}-m_{s}A'_{s})\times \mbox{\hspace{130pt}}\nn\\
(m_{q}A'_{q}+\frac{1}{2}m_{s}A'_{s}-\frac{3}{4}A'_{G})e^{-m^{2}_{\eta'}/M^{2}}\mbox{\hspace{70pt}}\nn\\
  =\frac{3}{\sqrt{2}\pi^{^{2}}}\left(\frac{\alpha_{s}}{\pi}\right)^{2} m_{s}^{2}M^{2}\left(\frac{7}{4}-\frac{1}{2}ln\frac{M^{2}}{\mu^{2}} \right)E_{0}\left(\frac{W^{2}}{M^{2}}\right)
\mbox{\hspace{10pt}}\nonumber\\
  -\frac{1}{\sqrt{2}\pi^{^{2}}}m_{s}^{2}M^{2}\left\{1+\frac{\alpha_{s}}{\pi}\left(\frac{17}{3}+2\gamma-2ln\frac{M^{2}}{\mu^{2}} \right)\right\}E_{0}\left(\frac{W^{2}}{M^{2}}\right)\nonumber\\
+\frac{4\sqrt{2}}{3}m_{s}\langle\bar{s}s\rangle-2\sqrt{2}\left(\frac{\alpha_{s}}{\pi}\right)^{2}m_{s}\langle\bar{s}s\rangle
\left(\gamma-ln\frac{M^{2}}{\mu^{2}}\right)\nonumber\\
+\frac{\sqrt{2}}{3}\frac{m^{2}_{s}}{M^{2}}  \left\langle\frac{\alpha_{s}}{\pi}G^{2}\right\rangle
+\frac{1}{\sqrt{2}}\frac{\alpha_{s}}{\pi}\frac{m^{2}_{s}}{M^{2}}\left\langle\frac{\alpha_{s}}{\pi}G^{2}\right\rangle
\left(1-\gamma+ln\frac{M^{2}}{\mu^{2}}\right)\nonumber\\
-\frac{8\sqrt{2}}{3}m^{3}_{s}\frac{\langle\bar{s}s\rangle}{M^{2}}-\sqrt{2}\frac{\alpha_{s}}{\pi}\frac{m_{s}}{M^{2}}
\langle\bar{s}g_{s}\sigma.G s\rangle\mbox{\hspace{47pt}}\nonumber\\
+\frac{32\sqrt{2}}{9}\pi^{2}\frac{\alpha_{s}}{\pi}\frac{m^{2}_{s}}{M^{4}}\kappa_{s}\langle\bar{s}s\rangle^{2}+\frac{2\sqrt{2}}{3}
\frac{m^{3}_{s}}{M^{4}}\langle\bar{s}g_{s}\sigma.G s\rangle\nonumber\\
  -\frac{3}{\sqrt{2}\pi^{^{2}}}\left(\frac{\alpha_{s}}{\pi}\right)^{2} m_{q}^{2}M^{2}\left(\frac{7}{4}-\frac{1}{2}ln\frac{M^{2}}{\mu^{2}} \right)E_{0}\left(\frac{W^{2}}{M^{2}}\right)
\mbox{\hspace{10pt}}\nonumber\\
  +\frac{1}{\sqrt{2}\pi^{^{2}}}m_{q}^{2}M^{2}\left\{1+\frac{\alpha_{s}}{\pi}\left(\frac{17}{3}+2\gamma-2ln\frac{M^{2}}{\mu^{2}} \right)\right\}E_{0}\left(\frac{W^{2}}{M^{2}}\right)\nonumber\\
  -\frac{4\sqrt{2}}{3}m_{q}\langle\bar{q}q\rangle +2\sqrt{2}\left(\frac{\alpha_{s}}{\pi}\right)^{2}m_{q}\langle\bar{q}q\rangle
\left(\gamma-ln\frac{M^{2}}{\mu^{2}}\right)\nonumber\\
-\frac{\sqrt{2}}{3}\frac{m^{2}_{q}}{M^{2}}  \left\langle\frac{\alpha_{s}}{\pi}G^{2}\right\rangle
-\frac{1}{\sqrt{2}}\frac{\alpha_{s}}{\pi}\frac{m^{2}_{q}}{M^{2}}\left\langle\frac{\alpha_{s}}{\pi}G^{2}\right\rangle
\left(1-\gamma+ln\frac{M^{2}}{\mu^{2}}\right)\nonumber\\
+\sqrt{2}\frac{\alpha_{s}}{\pi}\frac{m_{q}}{M^{2}}
\langle\bar{q}g_{s}\sigma.G q\rangle
-\frac{32\sqrt{2}}{9}\pi^{2}\frac{\alpha_{s}}{\pi}\frac{m^{2}_{q}}{M^{4}}
\kappa\langle\bar{q}q\rangle^{2} \mbox{\hspace{50pt}}
\eea
    With the constants as given by Eq.(20), we have fitted Eq.(26) in the range $ 1.0\; GeV^{2} < M^{2} < 1.9\; GeV^{2} $ and displayed it in Fig.3. The upper end of the range of $ M^{2} $ for this fit is kept somewhat higher than that for the fit of $ P_{L}^{88} $ because the required fit is better on the upper end side and not so well on the lower end side, as is clear from a comparison of Fig.3 with Fig. 1. In the next section, we have analyzed the effect of variation of the range of $ M^{2} $ over which fitting has been carried out. In this case also, the contribution of highest dimensional operator has been found to be less than 1\% while the resonance and continuum contribution is less than 25\% to the OPE side result. The quality of fit is somewhat poor: $ \chi=1.3\times10^{-3}$. From the fit we get
    \bea
    \frac{8\sqrt{2}}{3m^{2}_{\eta}}(m_{q}A_{q}-m_{s}A_{s})(m_{q}A_{q}+\frac{1}{2}m_{s}A_{s}-\frac{3}{4}A_{G})\nonumber\\
    = 4.64\times10^{-4}\; GeV^{4},\nn\\
    \frac{8\sqrt{2}}{3m^{2}_{\eta'}}(m_{q}A'_{q}-m_{s}A'_{s})(m_{q}A'_{q}+\frac{1}{2}m_{s}A'_{s}-\frac{3}{4}A'_{G})\nonumber\\
    = -6.414\times10^{-3}\; GeV^{4},\nn\\
    K_{08} = -3.152\times10^{-3}\; GeV^{4}\mbox{\hspace{10pt}}
    \eea
$ P_{L}^{88}$ and $ P_{L}^{08}$ provide two independent sum rules in which the phenomenological side of $ P_{L}^{08}$ contains two additional parameters $ A_{G} $ and  $ A'_{G} $ not present in $ P_{L}^{88}$. In both the sum rules the same parameters including condensates and continuum threshold have been used although the range of the fit of $ M^{2} $ is slightly larger for $ P_{L}^{08}$. Once the four parameters, namely, $ m_{q}A_{q}, m_{q}A'_{q}, m_{s}A_{s} $ and $ m_{s}A'_{s} $ have been determined from  $ P_{L}^{88}$ sum rule, this information can be fed in $ P_{L}^{08}$ sum rule to determine $ A_{G} $ and  $ A'_{G} $. From Eqs. (25) and (27), we get
\be
A_{G}=-0.02197\; GeV^{3},\;\;A'_{G}=-0.03704 \; GeV^{3}
\ee
Eq. (26) with $ m_{q}=0 $ on both sides of the equation has also been analyzed in a similar way and the fit has been displayed in Fig.4. For this fit $ \chi=1.0\times10^{-3}$. This fit gives
\bea
    \frac{8\sqrt{2}}{3\widetilde{m}^{2}_{\eta}}m_{s}A_{s}(\frac{1}{2}m_{s}A_{s}-\frac{3}{4}A_{G})\mbox{\hspace{5pt}}
    &=& -7.3\times10^{-3}\; GeV^{4},\nn\\
    \frac{8\sqrt{2}}{3\widetilde{m}^{2}_{\eta'}}m_{s}A'_{s}(\frac{1}{2}m_{s}A'_{s}-\frac{3}{4}A'_{G})\mbox{\hspace{5pt}}
    &=& 6.69\times10^{-3} \;GeV^{4},\nn\\
    \widetilde{K}_{08}\mbox{\hspace{5pt}} &=& -3.33\times10^{-3}\; GeV^{4}\mbox{\hspace{20pt}}
    \eea
    On substituting $ m_{s}A_{s} $ and  $ m_{s}A'_{s} $  from Eq. (25) we get
    \be
A_{G}=-0.02199\; GeV^{3},\;\;A'_{G}=-0.03616\; GeV^{3}
\ee
This shows that while  $ A_{G} $ remains practically unchanged,   $ |A'_{G}| $  decreases by $ \approx $ 2.4\%  from values given by Eq. (28). We will check the sensitivity of our result on $ m_{s}$ later.  We also find that Eq. (27) gives
\be
f_{0}=105.7 \;MeV,\;\; \theta_{0}=-5.3^{\circ}
\ee
while Eq. (29) gives
\be
f_{0}=110.0\; MeV,\;\; \theta_{0}=-8.1^{\circ}
\ee
compared to $f_{0} $ =142.3 MeV and $ \theta_{0}  = -11.1^{\circ} $ [12].  Part of the reason for this significant change is $ \widetilde{m}_{\eta} $ and $ \widetilde{m}_{\eta'} $ as against $ m_{\eta} $ and $ m_{\eta} $ used in Ref. [12] and different $ \kappa $ and  $ \kappa_{s} $ used in this work, and part of the reason is a small mistake in numerical evaluation of mixed condensate term in our previous work [12].

\section{ANALYSIS AND DISCUSSION}
In order to show that our results are sufficiently sensitive to distinguish the cases with $ m_{q}\neq0 $ from those with $ m_{q}=0 $, we have displayed the two curves  (with $ m_{q} $=5 MeV and $ m_{q}=0 $) on the same plot in Fig. 5 for BT $\left[-P_{L}^{88}\right] $ and in Fig. 6 for  BT $\left[-P_{L}^{08}\right] $ where BT stands for Borel transform.

      We have also checked the sensitivities of our results to variations of the parameters used in the computation. We have shown the plots of  BT $\left[-P_{L}^{88}\right] $ in Fig.7 as $ \langle\bar{q}q\rangle $ is varied by 20\% and $ \left\langle\frac{\alpha_{s}}{\pi}G^{2}\right\rangle $ is varied by 40\%, in Fig. 8 as both $ \alpha_{s} $ and $ m_{s} $ are varied by 10\% each, and in Fig. 9 as the continuum threshold $ W^{2} $ is varied by ±0.1 $ GeV^{2} $. Plots of  BT $\left[-P_{L}^{08}\right] $ are shown in Figs. 10, 11 and 12 respectively for similar variations of parameters. The quark pseudoscalar densities and anomaly matrix elements for $ \eta $ and $ \eta' $ mesons for each one of these variations have been determined separately. We have also determined these matrix elements when the range of the Borel mass squared,$ M^{2} $, over which fitting is carried out, is changed by ±0.1 $ GeV^{2} $ for both cases of BT $\left[-P_{L}^{88}\right] $ and  BT $\left[-P_{L}^{08}\right] $. $ \kappa $ and  $ \kappa_{s} $ appear in the matrix elements of the operators with highest dimension used whose contributions to the sum rule happens to be less than 1\%. Hence the error due to the uncertainties in their values will be small. The errors arising due to the uncertainties in numerical values of $ \kappa $ ,  $ \kappa_{s} $ as well as  $ \widetilde{m}_{\eta} $ and $ \widetilde{m}_{\eta'} $  being small, have been neglected. Our final result of this analysis is as follows:

      \bea
      m_{q}A_{q}\times10^{4}(GeV^{3})=7.45^{+0.75}_{-0.69}(m_{s})^{+0.14}_{-0.10}(\alpha_{s})^{+0.05}_{-0.06}(\langle\bar{q}q\rangle)
      \nn\\^{+0.08}_{+0.02} \left(\left\langle\frac{\alpha_{s}}{\pi}G^{2}\right\rangle\right)^{+0.65}_{-0.51}(W^{2})
      ^{+0.37}_{-0.39}(M^{2}),\nn\\
       m_{q}A'_{q}\times10^{4}(GeV^{3})=5.53^{+0.57}_{-0.58}(m_{s})^{+0.72}_{-0.63}(\alpha_{s})^{+0.70}_{-0.60}(\langle\bar{q}q\rangle)
      \nn\\^{+0.29}_{-0.63} \left(\left\langle\frac{\alpha_{s}}{\pi}G^{2}\right\rangle\right)^{+0.96}_{-0.74}(W^{2})
      ^{+1.12}_{-0.79}(M^{2}),\nn\\
      m_{s}A_{s}\times10^{2}(GeV^{3})=-2.841^{-0.286}_{+0.284}(m_{s})^{-0.209}_{+0.225}(\alpha_{s})^{-0.171}_{+0.151}(\langle\bar{q}q\rangle)
      \nn\\^{-0.094}_{+0.101} \left(\left\langle\frac{\alpha_{s}}{\pi}G^{2}\right\rangle\right)^{-0.211}_{+0.245}(W^{2})
      ^{-0.305}_{+0.265}(M^{2}),\nn\\
      m_{s}A'_{s}\times10^{2}(GeV^{3})=3.786^{+0.379}_{-0.378}(m_{s})^{+0.143}_{-0.149}(\alpha_{s})^{+0.104}_{-0.136}(\langle\bar{q}q\rangle)
      \nn\\^{+0.036}_{-0.037} \left(\left\langle\frac{\alpha_{s}}{\pi}G^{2}\right\rangle\right)^{+0.427}_{-0.464}(W^{2})
      ^{+0.299}_{-0.402}(M^{2}),\nn\\
      A_{G}\times10^{2}(GeV^{3})=-2.197^{-0.201}_{+0.244}(m_{s})^{-0.064}_{+0.043}(\alpha_{s})^{-0.062}_{+0.075}(\langle\bar{q}q\rangle)
      \nn\\^{-0.478}_{+0.280} \left(\left\langle\frac{\alpha_{s}}{\pi}G^{2}\right\rangle\right)^{-0.022}_{+0.258}(W^{2})
      ^{-0.127}_{-0.044}(M^{2}),\nn\\
      A'_{G}\times10^{2}(GeV^{3})=-3.704^{-0.251}_{+0.247}(m_{s})^{-0.624}_{+0.581}(\alpha_{s})^{-0.568}_{+0.299}(\langle\bar{q}q\rangle)
      \nn\\^{-0.400}_{+0.271} \left(\left\langle\frac{\alpha_{s}}{\pi}G^{2}\right\rangle\right)^{-0.148}_{+0.539}(W^{2})
      ^{-0.400}_{+0.130}(M^{2}).\nn\\
      \eea

      In Table I, we compare our results with those obtained by other authors. We give the total theoretical errors in our results of the matrix elements by adding the six individual theoretical errors, as given in Eq. (32),  in quadrature. The maximum error  $ \sim 34 $  \% is for $ m_{q}A'_{q} $ while the minimum error $ \sim 13 $  \% is for $ m_{q}A_{q} $.\\

      TABLE I : Comparison of our results on pseudoscalar densities and anomaly matrix elements of $ \eta $ and $ \eta' $ mesons with those obtained by other authors (for Ref.[3], numerical evaluation was done by us). \\
\scalebox{0.65}{
\begin{tabular} {|l|c|c|c|c|c|c|}
\hline
    Ref. & $ m_{q}A_{q}\times $ & $ m_{q}A'_{q}\times $ & $ m_{s}A_{s}\times $ &  $ m_{s}A'_{s}\times $ & $ A_{G} \times 10^{2}$ & $ A'_{G}\times10^{2} $ \\
     & $ 10^{4}(GeV^{3}) $ & $ 10^{4}(GeV^{3}) $ & $ 10^{2}(GeV^{3}) $ & $ 10^{2}(GeV^{3}) $ & $ (GeV^{3}) $ & $ (GeV^{3}) $\\\hline
This work & $ 7.45^{+1.07}_{-0.95} $ & $ 5.53^{+1.90}_{-1.63} $ & $ -2.84^{-0.55}_{+0.54}$ & $ 3.79^{+0.67}_{-0.75} $ & $ -2.20^{-0.54}_{+0.46} $ & $ -3.70^{-1.06}_{+0.93} $ \\\hline
Gerard & 6.25 & 5.46 & $(-2.7\pm0.4) $ & $ 5.45\pm 0.8 $ & &  \\ \& Kou [5] & & & $ (-2.9) $ & (3.45) & &  \\\hline
Pham [3] & & & $ -2.52 $ & $ 3.63 $ & & \\\hline
Cheng et & & & & & $-2.56 $ & $ -5.4 $  \\ al. [9] & & & & & $(-2.80) $ & $ (-5.7) $ \\\hline
Novikov et & & & & & $ -2.1 $ & $ -3.5 $ \\ al. [19] & & & & & & \\\hline
AG [6] & & & $ -2.85 $ & 3.55 & & \\\hline
BN [4] & 3.54 $ \pm10.62 $ & 3.54 $ \pm7.08 $ & $-2.75 $ & 3.40 & $-2.2 \pm 0.2 $ & $-5.7 \pm 0.2 $ \\\hline
Feldmann [8] & 7.07 & 5.66 & $-2.65 $ & 3.25 & $-1.2 $ & $-2.9 $ \\\hline
\end{tabular}}\\[10 pt]

    While authors of Refs.[4,6,8] use $1/N_{C} $ improved chiral perturbation theory and FKS-scheme [8] for $ \eta $ - $ \eta' $ mixing, GK [5] have used low-energy effective theory of QCD in large $ N_{C} $ limit with one mixing angle scheme for $ \eta - \eta' $  in octet-singlet basis. Their results on matrix elements of pseudoscalar densities and anomaly are consistent with our results with error bars, except for $ A_{G} $ obtained by Feldmann [8] which is numerically smaller than our result. It may be pointed out that Feldmann has set up- and down-quark masses to zero for this derivation, which is equivalent to neglecting $ M^{2}_{\pi}$ compared to  $ M^{2}_{K}$ in his derivation.\\

       Pham[3] has used nonet symmetry for the matrix elements of the pseudoscalar densities in $ \eta $ and $ \eta' $ states to calculate $ A_{s} $ and  $ A'_{s} $ by extending the symmetry for masses of  pseudoscalar nonets that includes the effect of $ U(1)_{A} $ QCD anomaly. Results on $ A_{G} $ and $ A'_{G} $ obtained by Novikov et al [19] is based on $ SU(3)_{flavor}$ symmetry and QCD sum rule in the limit of $ m_{q} $=0 (q=u,d). These results for both pairs of the matrix elements agree with our results.\\

         Cheng et al. [9] have introduced a new element, neither used by other authors referenced in Table-I nor in the present work, in form of $ \eta $ - $ \eta' $-G mixing, where G is the pseudoscalar glueball. The present work, based on two-angle scheme for $ \eta $ - $ \eta' $ mixing in octet-singlet basis and on QCD sum rule approach, is also free of contamination of higher mass states due to explicit use of continuum threshold. In addition, we have retained the three light quark masses to nonzero.\\

          For many applications it is convenient to introduce the quark-flavor basis states:
\be
|\eta_{q}\rangle=\frac{1}{\sqrt{2}}|u\bar{u}+d\bar{d}\rangle,\;\;|\eta_{s}\rangle=|s\bar{s}\rangle
\ee
The physical states $ \eta $ and $ \eta' $ are related to the flavor states through a unitary matrix $ U(\varphi)$ [1,2,4]:

\be
\left( \begin{array}{clrr} %
|\eta\rangle \\
|\eta'\rangle
\end{array} \right)
=U(\varphi)\left( \begin{array}{clrr} %
|\eta_{q}\rangle \\
|\eta_{s}\rangle
\end{array} \right)
\ee
where
\be
U(\varphi)=\left( \begin{array}{clrr} %
        \cos\varphi & -\sin \varphi \\
        \sin \varphi  & \cos\varphi
    \end{array} \right)
\ee

In the singlet-octet basis, where the angles $ \theta_{0}$ and $ \theta_{8}$ are small and their difference is comparable to these angles themselves, it is pertinent to keep them distinct. On the other hand, in the quark flavor basis
\be
\left| \frac{\varphi_{q}-\varphi_{s}}{\varphi_{q}+\varphi_{s}}  \right|\ll1
\ee
Hence, one generally deals with only one mixing angle $ \varphi\cong \varphi_{q}\cong \varphi_{s} $  in this basis for the sake of convenience [8].\\

In Table II, we have displayed the matrix elements of quark pseudoscalar densities and anomaly operators for quark flavor basis states for $ \varphi=39.3^{\circ} $ [1,4] using the central values of matrix elements given in Table I. We observe that OZI preserving and OZI violating matrix elements of pseudoscalar densities differ by an order of magnitude. This, again, indicates that the numerical values of $ A_{s} $ and $ A'_{s} $ will not be affected in a significant way when $ m_{q}$ is set to zero. Also notice that anomaly matrix element is larger for the state $ |\eta_{q}\rangle $ than $ |\eta_{s}\rangle $ , since it is energetically favorable for gluons to pair produce lighter quarks. \\

TABLE II : Values of our calculated matrix elements of quark pseudoscalar densities and anomaly operators (as given by central values in Table I) for quark flavor basis states.

\begin{tabular}{|l|c|r|}
\hline
    Matrix elements & Values of matrix elements \\ & ( $ GeV^{3} $ ) \\\hline
    $ \langle0|m_{u}\overline{u}i\gamma_{5}u+ m_{d}\overline{d}i\gamma_{5}d|\eta_{q}\rangle $ & 1.854 $ \times10^{-3} $ \\\hline
    $ \langle0|m_{u}\overline{u}i\gamma_{5}u+ m_{d}\overline{d}i\gamma_{5}d|\eta_{s}\rangle $ & $ - $ 0.88 $ \times10^{-4} $ \\\hline
    $ \langle0|m_{s}\overline{s}i\gamma_{5}s|\eta_{s}\rangle $ &  0.047317  \\\hline
    $ \langle0|m_{s}\overline{s}i\gamma_{5}s|\eta_{q}\rangle $ &  2.028 $ \times10^{-3} $ \\\hline
    $ \langle 0 |\frac{\alpha_{s}}{4\pi}G^{a}_{\mu\nu}\widetilde{G}^{a\mu\nu}|\eta_{q}\rangle $ & $ - $ 0.040460  \\\hline
    $ \langle 0 |\frac{\alpha_{s}}{4\pi}G^{a}_{\mu\nu}\widetilde{G}^{a\mu\nu}|\eta_{s}\rangle $ & $ - $ 0.014698  \\\hline

\end{tabular}\\[10 pt]

In summary, current-current correlators, using QCD sum rules give reasonable estimate for matrix elements of flavor-diagonal light quark pseudoscalar densities and axial anomaly operators between vacuum and $ \eta $ and $ \eta' $ meson states.\\

ACKNOWLEDGEMENT\\
Part of the work was done when the author was visiting University of Oregon, Eugene. The author thanks the authorities of the University for the hospitality.

\begin{figure}
\begin{center}
\scalebox{0.8}{\includegraphics{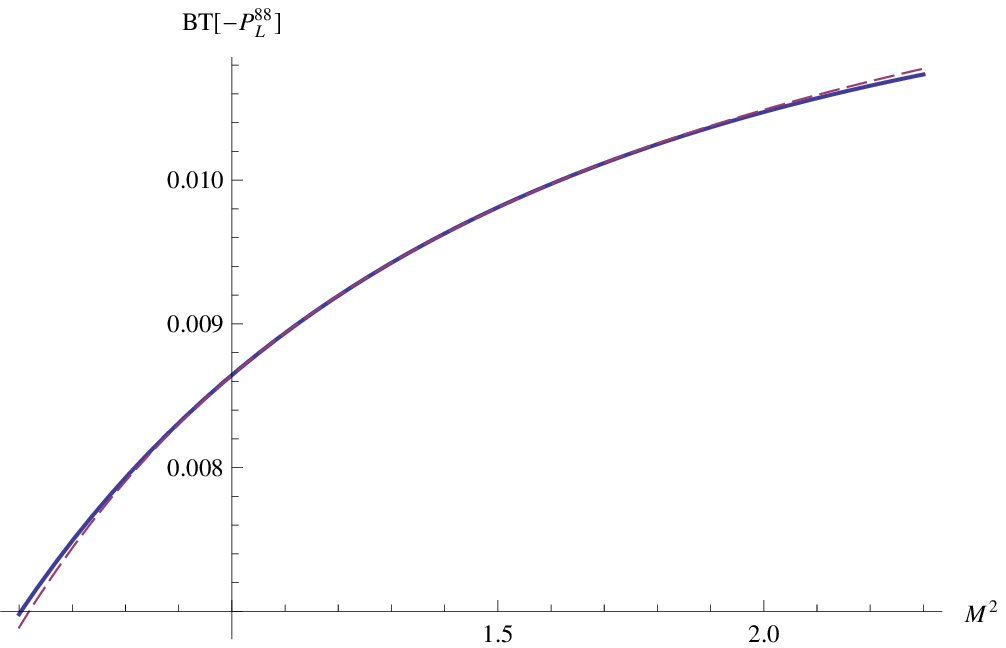}}
\caption{Plots of the two sides of  Eq. (18 ) (called $ BT[-P_{L}^{88}] $ ), as a function of $ M^{2} $ for  $ m_{q} $=5 MeV. The best fit in the region $ 1.0\; GeV^{2}\leq M^{2}\leq 1.7\; GeV^{2} $ corresponds to $ (f^{8}_{\eta})^{2} m^{2}_{\eta} =0.00758\; GeV^{4} $ and $ (f^{8}_{\eta'})^{2} m^{2}_{\eta'} =0.00404\; GeV^{4}. \chi=4.7\times10^{-5} $      (N=20) for the fit in the designated interval.}
\scalebox{0.8}{\includegraphics{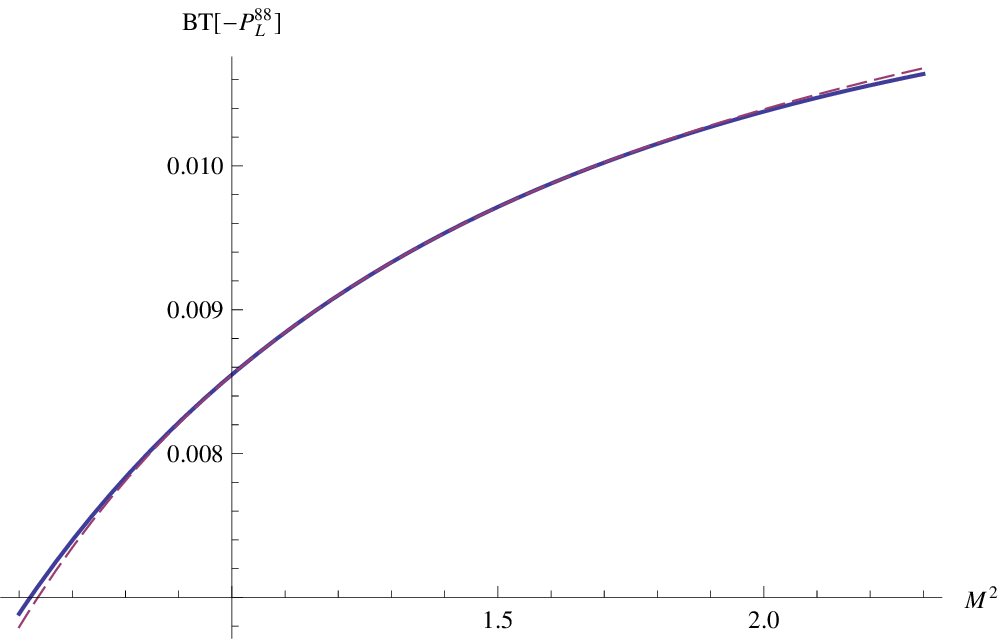}}
\caption{Plots of the two sides of  Eq. (19 ) (called $ BT[-P_{L}^{88}] $ ), as a function of $ M^{2} $ for  $ m_{q} $=0. The best fit in the region $ 1.0 \;GeV^{2}\leq M^{2}\leq 1.7 \;GeV^{2} $ corresponds to  $ (f^{8}_{\eta})^{2} \widetilde{m}^{2}_{\eta} =0.00752 \;GeV^{4} $  and  $ (f^{8}_{\eta'})^{2} \widetilde{m}^{2}_{\eta'} =0.00422 \;GeV^{4} $. The masses $ \widetilde{m}_{\eta} $ and $ \widetilde{m}_{\eta'} $ are the phenomenological masses obtained for $ m_{q} $=0 (see Eq.(17)). $ \chi=4.7\times10^{-5} $ for the fit in the designated interval. }
\scalebox{0.8}{\includegraphics{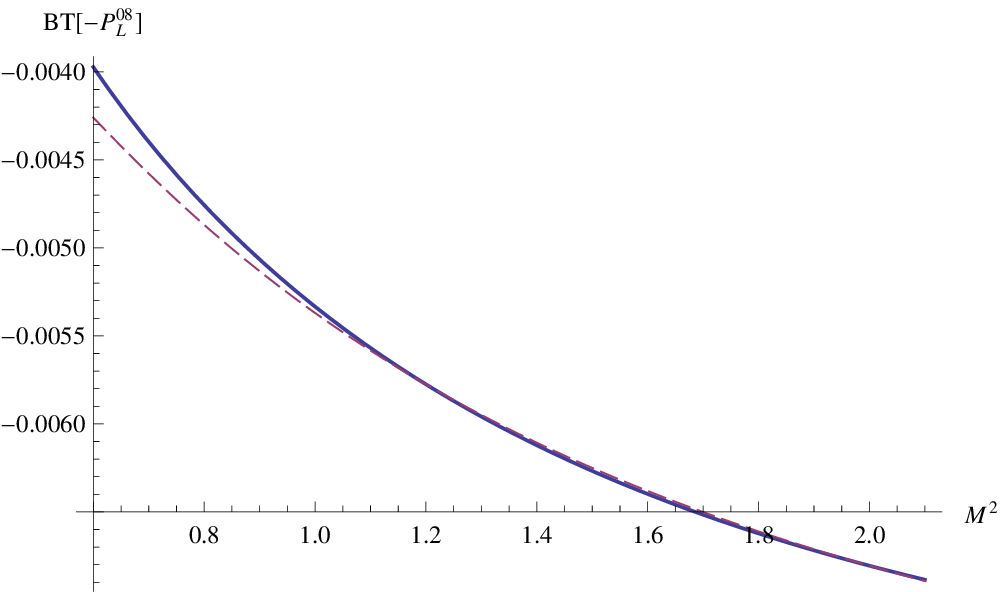}}
\caption{Plots of the two sides of  Eq. (26 ) (called $ BT[-P_{L}^{08}] $ ), as a function of $ M^{2} $ for  $ m_{q} $=5 MeV. The best fit in the region $ 1.0 \;GeV^{2}\leq M^{2}\leq 1.9 \;GeV^{2} $ corresponds to $ f^{0}_{\eta} f^{8}_{\eta} m^{2}_{\eta} =4.64\times10^{-4}\;GeV^{4} $ and $ f^{0}_{\eta'} f^{8}_{\eta'} m^{2}_{\eta'} =-6.414\times10^{-3}\; GeV^{4}. \chi=1.29\times10^{-3} $      (N=20) for the fit in the designated interval.}
\end{center}
\end{figure}

\begin{figure}
\begin{center}
\scalebox{0.8}{\includegraphics{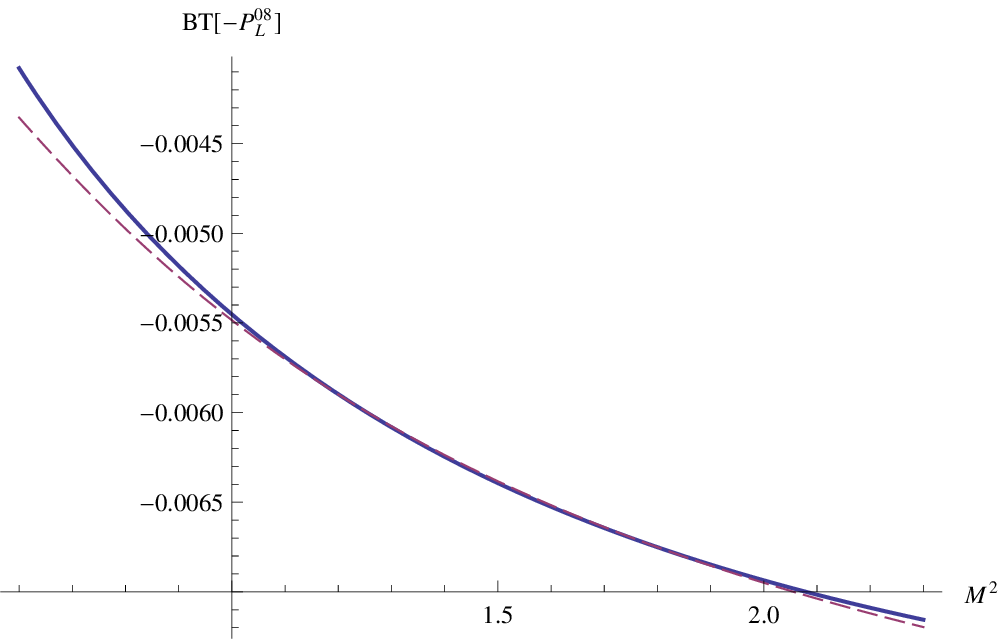}}
\caption{Plots of the two sides of  Eq. (26 ) (called $ BT[-P_{L}^{08}] $ ), as a function of $ M^{2} $ for  $ m_{q} $=0. The best fit in the region $ 1.0 \;GeV^{2}\leq M^{2}\leq 1.9 \;GeV^{2} $ corresponds to $ f^{0}_{\eta} f^{8}_{\eta} \widetilde{m}^{2}_{\eta} =7.3\times10^{-4}\;GeV^{4} $ and $ f^{0}_{\eta'} f^{8}_{\eta'} \widetilde{m}^{2}_{\eta'} =-6.689\times10^{-3}\; GeV^{4}$. The masses $ \widetilde{m}_{\eta} $ and $ \widetilde{m}_{\eta'} $ are the phenomenological masses obtained for $ m_{q} $=0 (see Eq.(17)). $ \chi=9.97\times10^{-4} $ (N=20) for the fit in the designated interval.}
\scalebox{0.8}{\includegraphics{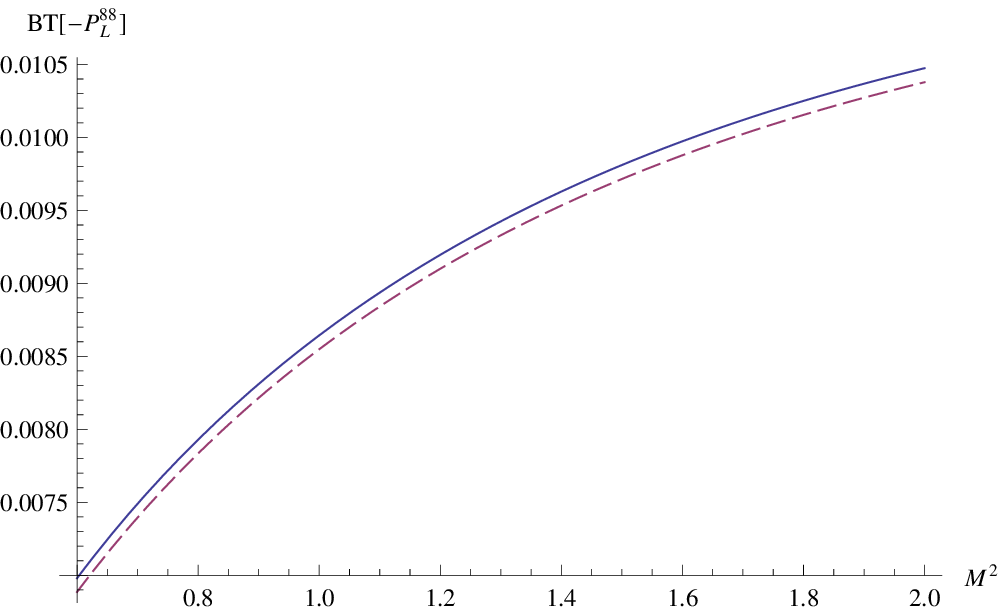}}
\caption{Plots of the two forms of  $ BT[-P_{L}^{88}] $, as a function of $ M^{2} $  for $ m_{q} $=5 MeV (solid line) and for $ m_{q} $=0 (dashed line).}
\scalebox{0.8}{\includegraphics{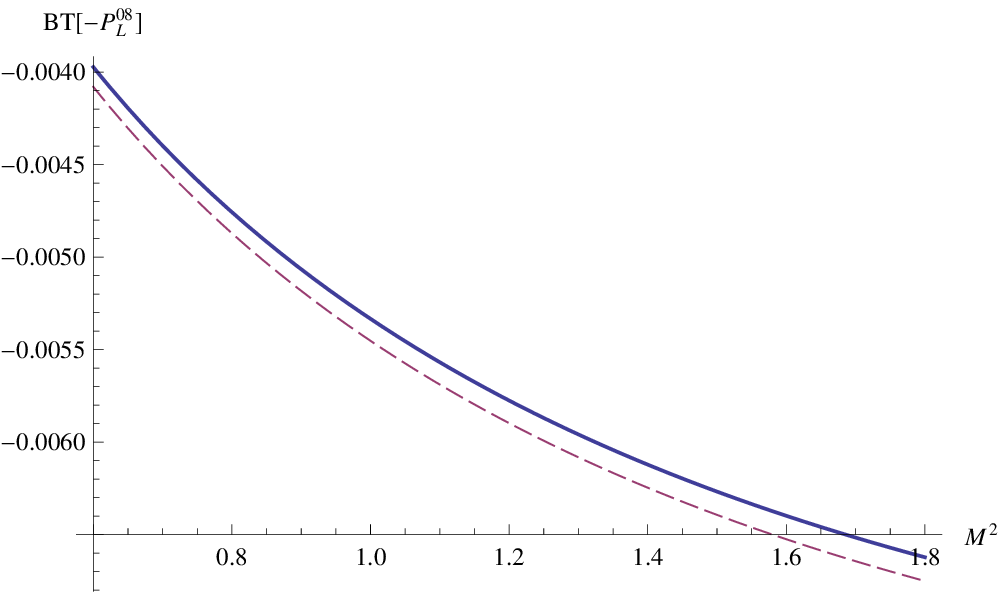}}
\caption{Plots of the two forms of  $ BT[-P_{L}^{08}] $, as a function of $ M^{2} $  for $ m_{q} $=5 MeV (solid line) and for $ m_{q} $=0 (dashed line).}
\end{center}
\end{figure}

\begin{figure}
\scalebox{0.5}{\includegraphics{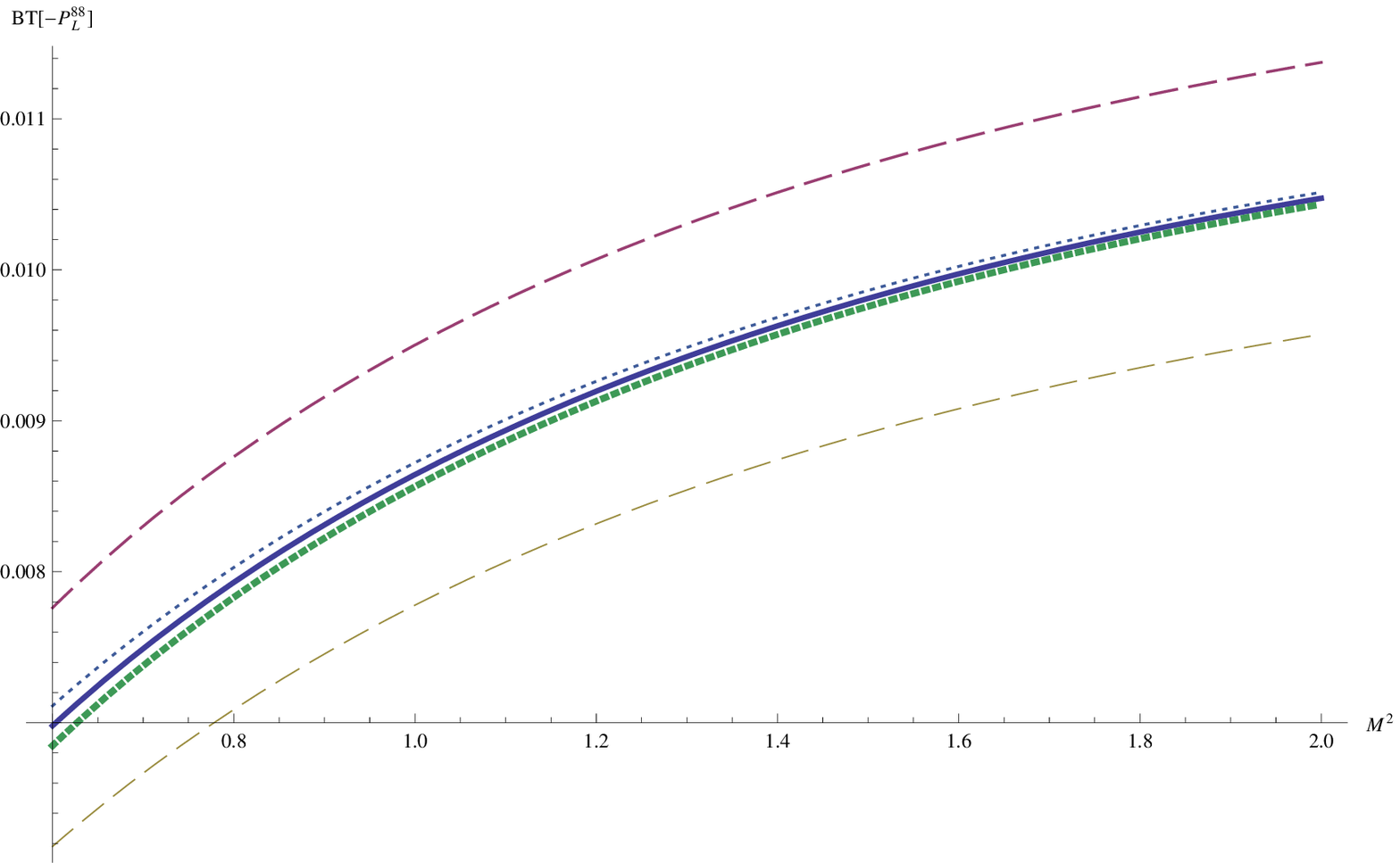}}
\caption{Plots of $ BT[-P_{L}^{88}] $  as a function of $ M^{2} $  for normal (solid), quark condensate changed by 20\% (dashed) and gluon condensate changed by 40\% (dotted).}
\scalebox{0.8}{\includegraphics{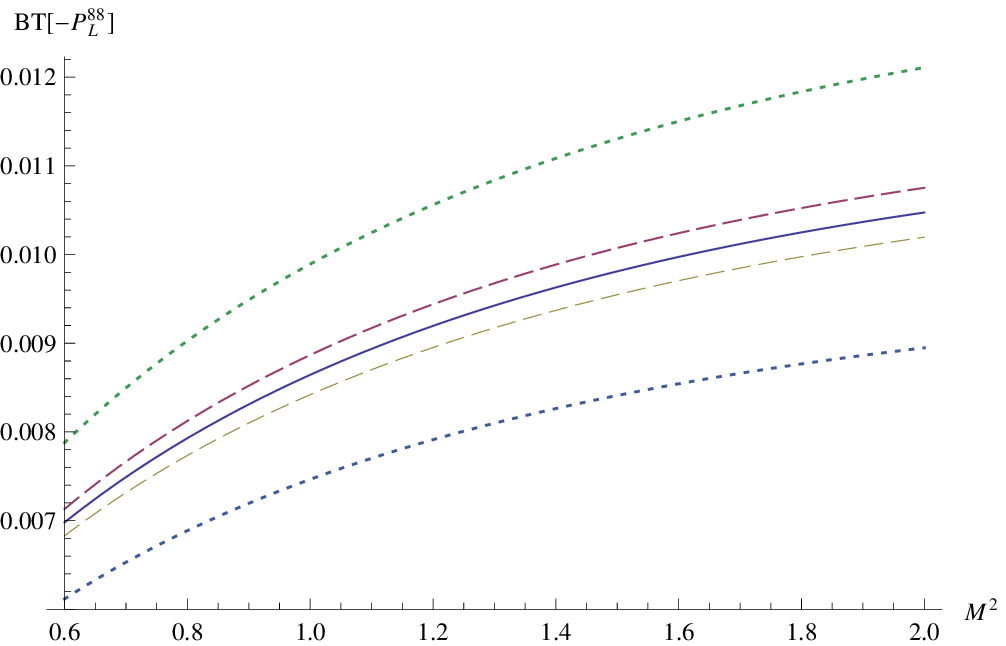}}
\caption{Plots of $ BT[-P_{L}^{88}] $  as a function of $ M^{2} $  for normal (solid), $ \alpha_{s} $ changed by 10\% (dashed), $ m_{s} $ changed by 10\% (dotted).}
\scalebox{0.8}{\includegraphics{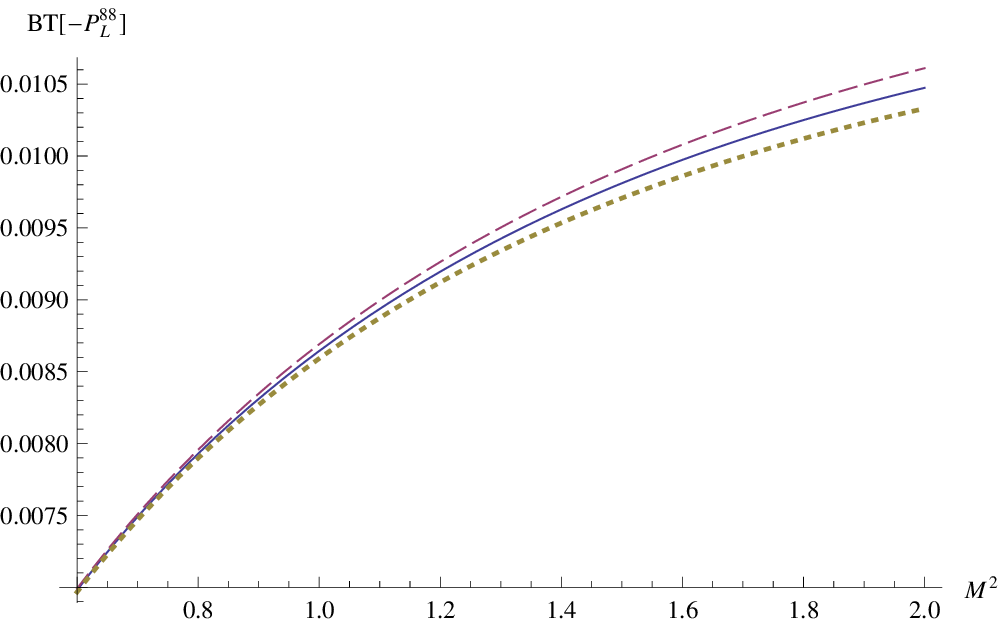}}
\caption{Plots of $ BT[-P_{L}^{88}] $  as a function of $ M^{2} $  for  $ W^{2}=2.3\; GeV^{2} $ (solid line), for  $ W^{2}=2.4 \;GeV^{2} $ (dashed line) and for $ W^{2}=2.2\; GeV^{2} $ (dotted line).}
\end{figure}

\begin{figure}
\scalebox{0.8}{\includegraphics{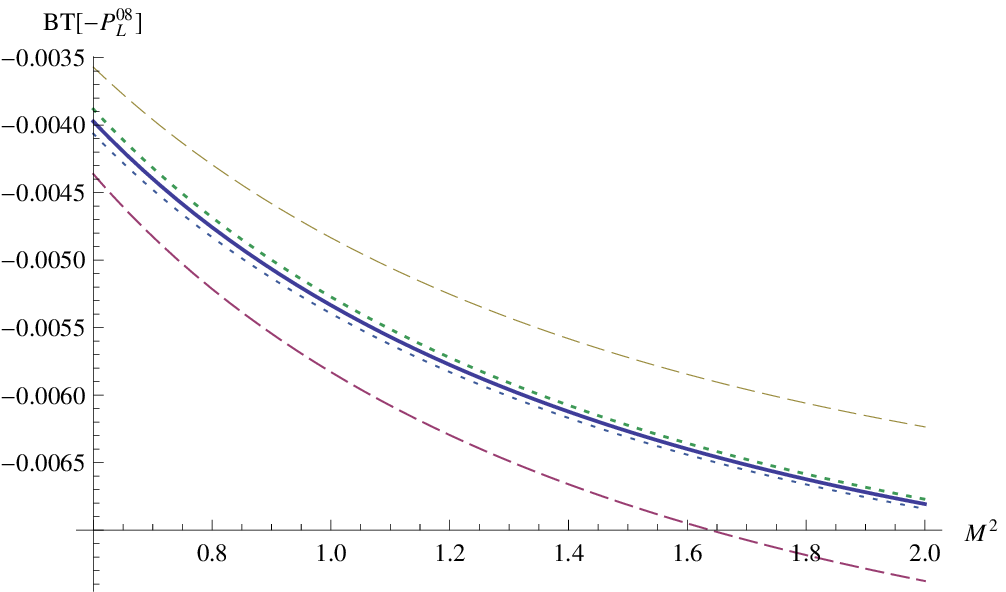}}
\caption{Plots of $ BT[-P_{L}^{08}] $  as a function of $ M^{2} $  for normal (solid), quark condensate changed by 20\% (dashed) and gluon condensate changed by 40\% (dotted).}
\scalebox{0.8}{\includegraphics{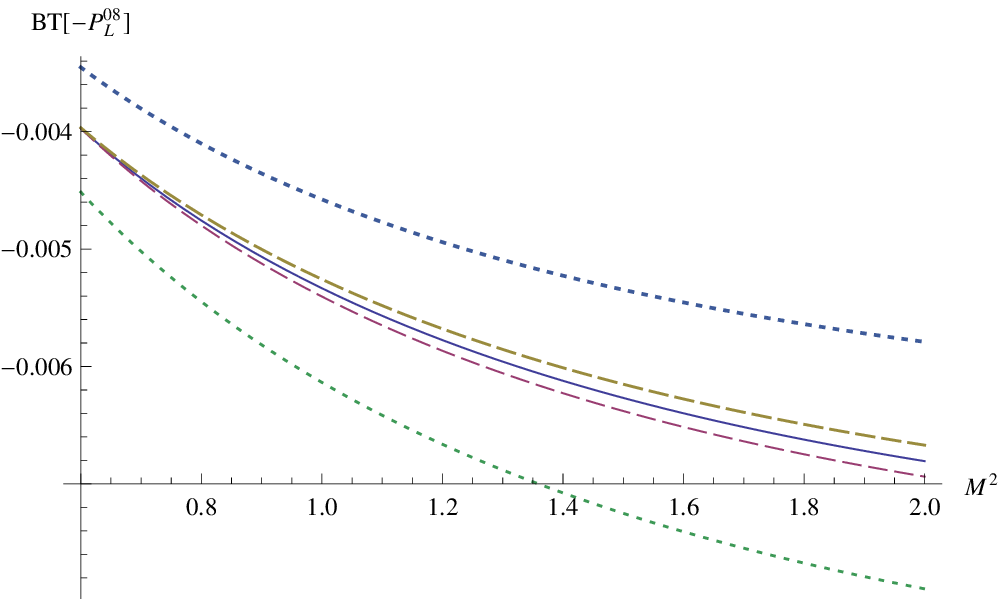}}
\caption{Plots of $ BT[-P_{L}^{08}] $  as a function of $ M^{2} $  for normal (solid), $ \alpha_{s} $ changed by 10\% (dashed), $ m_{s} $ changed by 10\% (dotted).}
\begin{flushright}
\scalebox{0.8}{\includegraphics{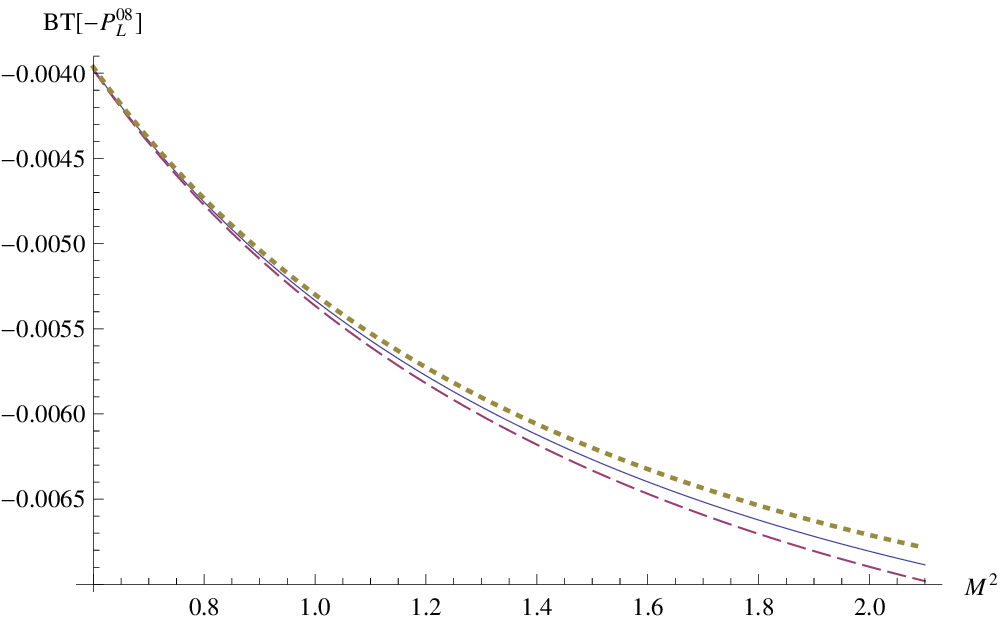}}
\caption{Plots of $ BT[-P_{L}^{08}] $  as a function of $ M^{2} $  for  $ W^{2}=2.3 \;GeV^{2} $ (solid line), for  $ W^{2}=2.4 \;GeV^{2} $ (dashed line) and for $ W^{2}=2.2\; GeV^{2} $ (dotted line).}
\end{flushright}
\end{figure}


\begin{thebibliography}{19}

\bibitem{charng}  Y.-Y. Charng, T. Kurimoto and H.-n. Li, Phys. Rev. D 74, 074024 (2006).\\
\bibitem{hsu} J.-F. Hsu, Y.-Y. Charng and H.-n. Li, arXiv:0711.4987 [hep-ph].\\
\bibitem{pham} T. N. Pham, Phys. Rev. D 77, 014024 (2008).\\
\bibitem{BN} M. Beneke and M. Neubert, Nucl. Phys. B651, 225 (2003).\\
\bibitem{kou} J.-M. Gerard and E. Kou, Phys. Rev. Lett. 97, 261804 (2006).\\
\bibitem{AG} A. Ali and C. Greub, Phys. Rev. D 57, 2996 (1998) [arXiv: hep-ph/9707251]\\
\bibitem{chen_g} A. G. Akeroyd, C. H. Chen, and C. Q. Geng, Phys. Rev. D 75, 054003 (2007) [hep-ph/0701012].\\
\bibitem{feld} T. Feldmann, Int. J. Mod. Phys. A15, 159 (2000).\\
\bibitem{cheng} 9. H.-Y. Cheng, H.-n. Li and K.-F. Liu, Phys. Rev. D 79, 014024 (2009).\\
\bibitem{jpp} J. P. Singh and A. B. Patel, J. Phys. G39, 015006 (2012).\\
\bibitem{chen2} J.-W. Chen, H.-M. Tsai and K. C. Weng, Phys. Rev. D 73, 054010 (2006).\\
\bibitem{jpjp} J. P. Singh and J. Pasupathy, Phys. Rev. D 79, 116005 (2009).\\
\bibitem{bass} S. D. Bass, Physica Scripta T99, 96 (2002) [hep-ph/0210018];
      S. D. Bass, arXiv 0812.5047.\\
 \bibitem{ven} G. Veneziano, Nucl. Phys. B159, 213 (1979); E. Witten, Ann. Phys. 128, 363 (1980).\\
\bibitem{vec} P. Di Vecchia and  G. Veneziano, Nucl. Phys. B171, 253 (1980).\\
\bibitem{wein} S. Weinberg, The Quantum Theory of Fields, Cambridge Univ. Press (1996).\\
\bibitem{jpl} J. P. Singh and F. X. Lee, Phys. Rev. C 76, 065210 (2007).\\
\bibitem{ioffe} B. L. Ioffe, V. S. Fadin and L. N. Lipatov, Quantum Chromodynamics: Perturbative and
        Nonperturbative Aspects, Cambridge Univ. Press (2010).\\

\bibitem{nov} V. A. Novikov, M. A. Shifman, A. I. Vainshtein and V. I. Zakharov, Nucl. Phys. B165, 55 (1980).\\
\end{thebibliography}
\end{document}